\documentclass[%
 reprint,
 amsmath,amssymb,
 aps,
 prd,
]{revtex4-2}

\usepackage{graphicx}
\usepackage{float}

\usepackage{silence}
\usepackage{dcolumn}
\usepackage{bm}
\usepackage{xcolor}
\usepackage{hyperref}

\usepackage[normalem]{ulem} 

\begin{document}

\preprint{APS/123-QED}

\title{Rethinking Resonance Detectability during Binary Neutron Star Inspiral: Accurate Mismatch Computations for Low-lying Dynamical Tides}

\author{Alberto Revilla-Pe\~{n}a$^{1,2}$}
\email{arevilla@icc.ub.edu}

\author{Ruxandra Bondarescu$^{1,2}$}
\email{ruxandra@icc.ub.edu}

\author{Andrew P.~Lundgren$^{3,4}$}
\homepage{https://www.icrea.cat/community/icreas/33290/andrew-lundgren/}

\author{Jordi Miralda-Escud\'{e}$^{1,2,4}$}

\affiliation{$^{1}$Institut de Ci\`encies del Cosmos (ICCUB), Universitat de Barcelona (UB),
c. Mart\'i i Franqu\`es, 1, 08028 Barcelona, Spain}

\affiliation{$^{2}$Departament de F\'{\i}sica Qu\`antica i Astrof\'{\i}sica (FQA),
Universitat de Barcelona (UB), 08028 Barcelona, Spain}

\affiliation{$^{3}$Institut de F\'{\i}sica d’Altes Energies (IFAE),
Universitat Aut\`onoma de Barcelona, E-08193 Bellaterra (Barcelona), Spain}

\affiliation{$^{4}$Instituci\'{o} Catalana de Recerca i Estudis Avan\c{c}ats (ICREA),
08010 Barcelona, Spain}

\date{\today}

\begin{abstract}
We compute deviations from observed gravitational wave signals, where the amplitude of the signal is unchanged.  As an example, we consider the detectability of low lying dynamical tides in binary neutron star or neutron star black hole mergers. Tidal forces can excite  oscillatory modes of one or both of the stars in the binary when the orbital frequency of the binary system sweeps through the resonant mode frequency dissipating energy into the vibrational mode.
The orbital energy loss to the vibrational mode extracts energy from the orbital motion, advancing the time to merger. The inspiral then continues with an excess phase and a time advance. Both will cause a mismatch when fitting to a system that has not gone through the resonance. To resolve this effect, we compute the mismatch for current and planned detectors using both a quasi-analytical approach that relies on the computation of moment integrals and an optimized version of the standard numerical match function. We conclude that detectability can occur for time advances of the order of 1 ms with advanced LVK detectors for an excess energy-flux that is a few percent of the gravitational wave emission. Our results contrast with previous work, which model this effect solely as a phase shift of the waveform or by using the difference in the number of cycles induced by the resonant behavior. We show that tidal resonance effects primarily cause a time advance of the merger, rather than a phase difference, and that the single-frequency approximation commonly used in the literature significantly overestimates the detectability of this effect.

\end{abstract}

\maketitle


\section{\label{sec:intro}Introduction}

Gravitational wave (GW) interferometers are the most precise rulers ever built by humans. Since the first detection of GWs in 2015 \cite{Abbott2016GW150914}, the sensitivity of the Advanced LIGO detectors has more than doubled and continues to increase. Following the O4 observing run \cite{LVKO4,LIGOScientific:2025snk,OfekBirnholtz:2025vcb}, the LIGO-Virgo-KAGRA (LVK \cite{LVK, Aasi_2015, Acernese_2015}) collaboration is planning the O5 run, featuring the A+ upgrade \cite{Barsotti2018Aplus}, which is expected to further double the network’s detection volume and extend its reach to more distant and fainter sources. Beyond next LVK runs, third-generation observatories such as the Einstein Telescope (ET) \cite{ET} and Cosmic Explorer (CE) \cite{CE} aim to push the sensitivity up by another order of magnitude, probing compact-binary coalescences throughout the observable universe.

 GW observations offer new probes of the Universe pushing the frontiers of known compact object physics into the unknown. Coalescing neutron star - neutron star and neutron star - black hole (NS-NS and NS-BH) are particularly interesting because they can be multimessenger sources that leave signatures both through their gravitational wave \cite{abbott2017gw170817, abbott2020gw190425} and electromagnetic emissions \cite{abbott2017multi}. These observations constrain the internal structure of neutron stars and the physics of dense matter. They enabled the first constraints on the equation of state, radius and tidal-deformability of each NS \cite{abbott2017gw170817,abbott2018gw170817} from the gravitational waveform.  
 
 Current ground based GW detectors observe NS-NS and NS-BH binaries in the last few minutes of their lives, when their orbital frequency ranges over 10 to 1000 Hz. 
As these detectors become more accurate, additional constraints can be obtained.
Detailed observations of the late in-spiral phase, as well as the final GW front at the time of the merger, will provide valuable information on the equation of state and possible phase transitions of matter in neutron stars.
In these final minutes oscillation modes of neutron stars may resonate with the orbital frequency when the frequency of the tidal driving force equals the intrinsic mode frequency, resulting in energy transfer and angular momentum transfer from the binary orbit to individual neutron stars. This transfer affects the gravitational waveform primarily by advancing the merging time.

In the present study, we focus on modeling the imprint of resonant-mode excitation on the gravitational waveform. The response of the star to tidal excitation provides a window into interior neutron star physics, probing both its crust and core, which includes material up to many times the nuclear density in states inaccessible to laboratories on Earth \cite{graber2017neutron}. 

The tidal excitation of neutron star (NS) normal modes in a binary merger was first studied in detail by Lai in 1994 \cite{lai1994resonant}.  A number of modes can be excited when the orbital frequency sweeps through the 10-1000 Hz frequency band.  These include $\ell =2$ core g-modes, which couple to the tidal field leaving an imprint on the waveform. Its properties depend on the equation of state and on the symmetry energy of dense nuclear matter. Other modes that can become resonant include the $\ell =2 $ core-crust interface i-mode. First identified by \cite{mcdermott1988nonradial}, this excitation mode can fracture the crust through a large shear strain near the base of the crust \cite{tsang2013shattering,neill2021resonant,neill2024strengthening,neill2025resonant}. These can result in detectable flares \cite{tsang2012resonant} that may have been observed as precursors of the main flare in short Gamma-ray bursts (SGRBs) \cite{troja2010precursors,zhong2019precursors,coppin2020identification}.  More recently, it has been shown that compositional g-modes can also penetrate the crust and the concentrated strain at the base of the crust can contribute to the generation of cracks \cite{nbk7-8kts}. The coupling to the tidal field is comparable with that of the corresponding $\ell =2 $ i-mode \cite{nbk7-8kts, 10.1093/mnras/stab870, PhysRevD.107.083023}. Moreover, once stratification is included, the i-mode arising from the solid–liquid
interface at the base of the crust mixes with the g-mode and loses its identity
as a distinct mode \cite{nbk7-8kts}. New studies also suggest the presence of interface modes associated with first-order phase transitions, also called density-discontinuity g-modes \cite{8hvq-6dy7,k7l9-hw8g}. 
These modes can reach high values of tidal coupling, but were shown to exhibit high resonant frequencies $1-2~kHz$ \cite{8hvq-6dy7}. 
 
The strongest tidal coupling is exhibited by the f-mode, which can also perturb the crust, but its frequencies of  $\gtrsim 10^3$ Hz are in resonance beyond the LVK detectability window, and possibly not reached by the binary before merger. However, if an inspiralling NS is counter-rotating, the f-mode resonant frequency could be lower in the inertial frame reaching $\sim 800$ Hz \cite{Yu_2024,PhysRevResearch.3.033129}. Other modes like the r-modes couple only weakly to the tidal field \cite{FlanaganRacine} and are generally relevant for rapid rotators \cite{bondarescu2013nonlinear, bondarescu2009spinning, bondarescu2007spin}. Their detectability has been explored in the context of tidal resonances for third-generation interferometers \cite{gupta2022determiningequationstateneutron}. 
 
 In this paper, we perform a first parameter study that estimates the detectability  of tidal resonances for current and future detectors as a function of the mode frequency and the time advance of the merger (or equivanlently the energy transferred to the resonance). These results are agnostic to the values that are possible or likely due to interior neutron star physics.


Recent work by Counsell et al. \cite{Counsell:2024pua} has reevaluated the calculations of Lai 1994 \cite{lai1994resonant} for modern equations of state. They assess the detectability of resonant modes via an approximation proposed by Read \cite{read2023waveform}, which provides a rough bound but does not give an accurate answer.  Moreover, up to now phase differences between waveforms have been solely treated as phase offsets \cite{yu2017resonant,Edelman_2021,read2023waveform, ho2023new,Counsell:2024pua}, which are induced when orbital energy is rapidly lost and then regained, somehow, in a short time. However, in an astrophysical setting, the energy is transferred to a resonant mode and then dissipated through the star. It cannot be regained. This loss of orbital energy causes the stars to merge sooner, which may be detectable when comparing resonant and non-resonant waveforms.

To estimate the resonance detectability we compare the resonant waveform with the unperturbed waveform. In Sec.~\ref{sec:detect} we introduce the formalism necessary to detect these tidal resonances. We present the match-filtering techniques that are applied to extract relevant physical information from the detected signal. The mismatch is computed from the inner product between a waveform template including resonance effects and this same template without. An expansion around the unperturbed waveform up to quadratic order results in an approximation to the match that is valid for small resonances \cite{owen1996search, owen1999matched} and provides a fast and simple alternative to a numerical computation using the FFT. In this paper, we use the PyCBC software \cite{pyCBC,Allen:2005fk,Nitz:2017svb,DalCanton:2014hxh} for numerical computations of matched filtering.  In Sec.~\ref{sec:orb} we describe the binary system inspiral phase and the waveform emitted in the frequency domain. The energy-flux function is written to leading order, and modified in Sec.~\ref{sec:resmodel} to include the extra energy arising from the resonance. After the orbit has moved out of resonance, only the phase and end time have changed. Apart from this, the inspiral process proceeds normally. We assume the resonance is sharp and can be modeled as a delta function. 
Next, in Sec.~\ref{sec:dettech} we develop improved match-filtering techniques and adapt Lindblom indistinguishability criterion \cite{Lindblom_2008} for time shifts detection. 
Our results in Sec.~\ref{sec:detresults} show that previous detectability predictions provided by \cite{read2023waveform} overestimate the resonance effect by an order of magnitude. Sec.~\ref{sec:concl} concludes by summarizing the prospects of resonance detectability for various detectors including the LIGO detectors, the Einstein Telescope and Cosmic Explorer.

\section{\label{sec:detect}Detectability of Subtle Phase Shifts: methodology}


We begin by briefly reviewing the statistical methods for detecting and estimating the parameters of a GW signal in a noisy datastream. Detector data $d$ consist of detector noise $n$ plus possibly a signal $s$. 
\begin{equation}
    d(f) = n(f) + s(f) ~.
\end{equation}
We work in the frequency domain. The noise is assumed to be Gaussian and stationary, with a one-sided noise power spectral density $S_n(f)$. Only positive frequencies need to be considered because the signals in the time domain are real (so the negative frequencies are simply the complex conjugate of the corresponding positive frequencies). It is natural to define an inner product on the space of possible data as
\begin{equation}\label{eq:prodef}
\langle a, b \rangle = 2 \int
\frac{a^{*}(f) b(f) + a(f) b^{*}(f)}{S_n(f)}~ df ~,
\end{equation}
which is weighted inversely by the noise power. This inner product is a real quantity and is the fundamental quantity used in matched filtering.  The integral is in principle over all positive frequencies, but in practice the detectable power is contained between the minimum and maximum frequencies $f_{\rm min}$ and $f_{\rm max}$. All integrals will be taken to have these limits. 

A parametrized model for the signal is called a \textit{template}, e.g. $h(f; \lambda)$. The set of parameters $\lambda$ generally include the time and phase of the coalescence, masses and spins of the binary, and any other unknowns.
The coalescence time and phase of a signal are unknown a priori, so we must search over their possible values. The effect of a time shift $\Delta t$ and a phase shift $\Delta \phi$ in the frequency domain waveform is
\begin{equation}
    h(f; \Delta t, \Delta \phi) = h(f) e^{i (2 \pi f \Delta t + \Delta \phi)} ~.
\end{equation}
Calculations are done by generating the template with a fixed coalescence time and phase, then using the above factor to shift to any other time and phase values. The overall phase of the waveform is related to the initial orbital phase; shifting the orbital phase by some angle shifts the gravitational-wave phase by twice that amount. In the most general case, e.g. when the waveform is evolving quickly near merger or has multiple harmonics, a shift of orbital phase does not simply cause an overall phase shift in the frequency-domain waveform. However, during the inspiral and neglecting harmonics beyond the quadrupolar mode, the orbital phase is almost the same as an overall frequency-domain phase factor.

The SNR timeseries of a template $h$ with data $d$ is defined by
\begin{eqnarray}\label{eq:SNR_ts}
    \rho(t, \phi) &=& \frac{\langle h(f; t, \phi), d \rangle}{\sqrt{\langle h, h \rangle}} \\
    &=& \frac{1}{\sqrt{\langle h, h \rangle}} \mathrm{Re} \left[ 4 \int_{f_{\rm min}}^ {f_{\rm max}} \frac{h^*(f) d(f)}{S_n(f)} e^{-i (2 \pi f  t + \phi)} df \right]~. \nonumber
\end{eqnarray}
When the data contains the template, or a waveform sufficiently similar to the template, the SNR timeseries will have a peak. The time of the peak indicates where in time the template occurs within the data.
Calculating SNR timeseries is numerically accomplished with the inverse Fast Fourier Transform, so the search over time and phase can be accomplished very efficiently.

\subsection{Match}

The similarity of two waveforms is measured by the \emph{overlap} defined by
\begin{equation} \label{eq:overlap}
    {\cal O}(a, b) =  \frac{\langle a, b \rangle}{\sqrt{\langle a, a \rangle~\langle b, b \rangle}} ~.
\end{equation}
The overlap is insensitive to an overall amplitude scaling of either of the waveforms. Its maximum value is one, attained when $a$ and $b$ are identical up to an overall amplitude scaling (i.e. multiplication by a real number other than zero).

For any astrophysical GW signal, the true time and phase are unknown, which is why the full SNR timeseries must be computed. As we do not know the other parameters $\lambda$ of the signal, we need to compare with many possible templates. When computing the SNR using a template $a$ when the signal has a different waveform $b$, the time and phase of the peak will generally be shifted. The relevant measure of similarity of waveforms is the \emph{match}, the overlap maximized over time and phase offsets between two waveforms: 
\begin{eqnarray}
  \label{eq:match_definition}
    {\cal M}(a, b) &=&  \max \limits_{\Delta t,\Delta\phi} \frac{\langle a(f) , b(f; \Delta t, \Delta\phi) \rangle}{\sqrt{\langle a, a \rangle ~ \langle b, b \rangle}} ~.
 \end{eqnarray}

As with the SNR, the match function can be efficiently computed numerically by using the Fast Fourier Transform (FFT). However, the required number of sample points (the duration times sample rate) in the FFT can be quite large. The duration of the FFT should be at least as long as the time it takes for the template to evolve from $f_{\rm min}$ to $f_{\rm max}$, though this can be relaxed when the waveforms are very similar. The sample rate has two constraints. It must be at least twice $f_{\rm max}$ due to the Nyquist criterion, and it must also be sufficient to resolve the peak of the overlap sufficiently finely to perform an accurate computation of the match. In the worst case, the peak will fall exactly between two time samples, and the mismatch caused by this must be small compared to the mismatch due to the perturbation. When trying to find the mismatch due to small perturbations, it can be more efficient and more illuminating to use an approximation for the match.

\subsection{Quadratic Approximation}
\label{QA}
The matched filter is very sensitive to the frequency-dependent phase of the waveform, and much less sensitive to modifications of the amplitude. In addition, the physical effect we consider in this paper primarily affects the phase evolution of the waveform but not the amplitude. So we will only consider waveforms with perturbations to the phase. Starting from a base waveform $h_0$, we compare it to a perturbed waveform $h_\epsilon$ written as
\begin{equation}\label{eq:phasediff}
h_\epsilon(f) = h_0(f)~e^{i \epsilon \Delta \Psi(f) } 
\end{equation}

Our goal is now to approximately find the overlap between $h_0$ and $h_\epsilon$. We normalize $h_0$ to $\langle h_0, h_0\rangle = 1$, which also normalizes $h_\epsilon$, allowing the overlap to be written as
\begin{equation}
    {\cal O}(h_0, h_\epsilon) =  4 ~{\rm Re}\int \frac{|h_0(f)|^2}{S_n(f)} e^{i \epsilon\Delta\Psi(f)} d f~.
\end{equation}

We approximate the overlap by Taylor expanding the complex exponential terms to quadratic order in $\epsilon$. We have
\begin{eqnarray}
\label{eq:match_quadratic}
    {\cal O}(h_0, h_\epsilon) &=& 4~ {\rm Re} \int \frac{|h_0(f)|^2}{S_n(f)} \Big( 1 + i \epsilon \Delta\Psi(f)\nonumber\\
    &-&\frac{1}{2} \epsilon^2 \Delta\Psi(f)^2 \Big)~df~. 
\end{eqnarray}
The linear terms all vanish because $\langle i h, h \rangle = 0$ for any $h$ (see Eq.~\eqref{eq:prodef}). To express the other terms, we define noise moments of any function $\alpha(f)$ as
\begin{eqnarray}\label{eq:momentdef}
    I[\alpha(f)] &=& 4 \, {\rm Re} \int \frac{|h_0(f)|^2}{S_n(f)} \alpha(f)~\mathrm{d}f~, \\
    \mathcal{J}[\alpha(f)] &=& I[\alpha(f)]~/~I[1]~.
\end{eqnarray}
These are a generalization of \cite{owen1999matched}, where it was assumed that $|h_0|^2 \propto f^{-7/3}$, the leading-order post-Newtonian amplitude of an inspiral. The division by $I[1]$ in the definition of $\mathcal{J}$ comes from enforcing the normalization of $\langle h_0, h_0 \rangle = 1$; this makes $\mathcal{J}[1] \equiv 1$. Applying this to Eq.~\eqref{eq:match_quadratic}, we have
\begin{eqnarray}
{\cal O}(h_0, h_\epsilon) &=& 1 - \frac{1}{2} \epsilon^2 \mathcal{J}[\Delta\Psi^2]
\end{eqnarray}

To compute the match, we need to add phase and time offsets to the phase of one of the templates $h(f)\rightarrow  h(f;\delta\phi, \delta t)=h(f)e^{i( \delta\phi + 2\pi f \delta t) }$ to perform the maximization over those offsets
\begin{equation}\label{eq:matchdefquad}
    \mathcal{M}(h_0, h_\epsilon) = \max \limits_{\delta\phi,\delta t}~{\cal O}(h_0(f), h_\epsilon(f;\delta\phi,\delta t))~.
\end{equation}
For notational convenience, we will collect the waveform perturbation and time offset together in the term $\delta \psi = \epsilon\,\Delta\Psi + 2\pi f\,\delta t$.
The maximization over phase is equivalent to taking the absolute value
\begin{eqnarray}\label{eq:maxphifast}
\max \limits_{\delta\phi}~{\cal O}(h_0, h_\epsilon)&=&\max \limits_{\delta\phi}~4~ {\rm Re} \int \frac{|h_0|^2}{S_n} e^{i (\delta\psi + \delta\phi) } df \nonumber\\
&=&4~ {\rm Abs} \int \frac{|h_0|^2}{S_n} e^{i \delta \psi} df ~.
\end{eqnarray}
The result is
\begin{equation}\label{eq:Omaxphase}
\max \limits_{\delta\phi}{\cal O}(h_0, h_\epsilon)
= 1 - \frac{1}{2} \left( \mathcal{J}[\delta\psi^2] - \mathcal{J}[\delta\psi]^2 \right) ~.
\end{equation}

Now we can compute the match by maximizing Eq.~\eqref{eq:Omaxphase} over the time offset $\delta t$. We expand $\delta\psi$, yielding
\begin{eqnarray}
\max \limits_{\delta\phi}{\cal O}(h_0, h_\epsilon)
= 1 - \frac{1}{2} \Big( & & \mathcal{J}[(\epsilon\,\Delta\Psi+2\pi f\,\delta t)^2]\nonumber\\
& & - \mathcal{J}[\epsilon\,\Delta\Psi+2\pi f\,\delta t]^2 \Big)~.
\end{eqnarray}

This is quadratic in $\delta t$. Taking the maximum, we arrive at
\begin{equation}\label{eq:match_quad}
{\cal M}(h_0,h_\epsilon) = 1 - \frac{1}{2} g_{\Psi\Psi} \epsilon^2
\end{equation}
with
\begin{eqnarray}\label{eq:quadratic_coefficient}
g_{\Psi\Psi} & = &\mathcal{J}[\Delta\Psi^2] - \mathcal{J}[\Delta\Psi]^2\\
&-& \frac{(\mathcal{J}[(2\pi f) \Delta\Psi]-\mathcal{J}[2\pi f]\mathcal{J}[ \Delta\Psi])^2}{\mathcal{J}[(2\pi f)^2]-\mathcal{J}[2\pi f]^2} \nonumber ~.
\end{eqnarray}
The quantity $g_{\Psi\Psi}$ expresses how sensitive the match is to a small change in the perturbation $\epsilon$. It is fixed for a given detector PSD and given template $h_0$.

\section{\label{sec:orb} Binary System and GW emission}

The method above is quite general, but we now turn toward the motivating example for this paper: the detection of a neutron star resonant mode due to its excitation by tidal driving. Neutron stars possess a spectrum of vibrational modes depending on the restoring force \cite{Cowling1941, KokkotasSchmidtLRR}. 

The fundamental f-mode has high coupling to the tidal field, but its eigenfrequency is typically in the kHz range, which makes detection challenging. The $\ell = m = 2$ g-mode, which is restored by buoyancy, is particularly interesting because it couples with the tidal field and its frequency is in the hundreds of Hz range. Similar frequencies and higher couplings are achieved by the $\ell = m = 2$ crust-core interface modes.
Moreover, the frequency, damping and other characteristics vary depending on interior neutron star physics, like superfluidy \cite{yu2017resonant} and leptonic buoyancy \cite{10.1093/mnrasl/slu061,Rau_2018}. The importance of the latter effect was first realized by Kantor and Gusakov \cite{10.1093/mnrasl/slu061}. The presence of a quark matter core also has an impact on vibrational modes \citep[e.g.,][]{jaikumar2021g}. These different modes can increase the total rate of energy transfer from the orbit to the neutron star modes or to gravitational waves, which we refer to as energy-flux.


Our results will apply to either the NS-NS or NS-BH case; we will generically refer to either system as BNS. If the system has two neutron stars and they have similar resonant frequencies, then the effect simply doubles. If the stars have quite different resonant frequencies, or possess more than one significant resonance, then it is straightforward to extend the methods presented to that case.

\subsection{\label{sec:orbit}Orbital Equations}

Throughout what follows we use units where $G=c=1$. A solar mass corresponds to $G M_\odot/c^3 = 4.925 \times 10^{-6}$ sec.  Hence both $M f$ and $t / M$ are unitless. The binary can be parametrized in terms of the symmetric mass ratio $\eta = (m_1 m_2)/M^2,$ and the chirp mass ${\cal M} = M \ \eta^{3/5}$. Here $M = m_1 + m_2$ is the total mass and $m_1$ and $m_2$ are the component masses.
 
We assume the orbit is circular. Binaries are expected to circularize due to gravitational radiation emission well before reaching orbital frequencies accessible to current ground-based interferometers.  However, if the system is dynamically formed it may retain non-negligible eccentricity  \cite{PhysRevD.110.103043}. Such situations are left to future work. Far from the relativistic regime, a circular orbit of orbital separation $D$ has orbital velocity $v^2 \sim 1/D$.
The gravitational-wave frequency $f$ is twice the orbital frequency. It is related to $v$ according to Kepler's Law
\begin{equation}\label{eq:v=v(f)}
v = (\pi M f)^{1/3} ~.
\end{equation}
The inspiral is accurately described by a post-Newtonian expansion in the characteristic velocity $v$ until the tidal disruption and merger begins.

The orbit shrinks due to the loss of energy to gravitational waves. In the adiabatic approximation, the evolution is slow compared to the orbital period, and the radial velocity is negligible.  The orbital evolution is then determined by the energy balance equation
\begin{equation}
\label{energy_balance}
\frac{dE(v)}{dt} = - {\cal F}(v),
\end{equation}
where $E(v)$ is the orbital energy and ${\cal F}(v)$ is the gravitational-wave luminosity (which is also referred to as energy-flux or flux in some GW literature, e.g., \cite{BIOPS}).

The post-Newtonian expansions are Taylor series in the orbital velocity $v$, with logarithmic corrections appearing at higher orders. They are known to high PN order for the point-particle phasing and energy-flux relevant to compact-binary inspiral, while tidal contributions enter at higher order in $v$ in both the waveform and the energy-flux. In this paper, considering small resonance effects in the inspiral waveform, we only need to keep the leading order of $E(v)$ and ${\cal F}(v)$, which can be found in Eq.~\eqref{EF} and in Eq.~\eqref{dtdf} of \cite{BIOPS}:
\begin{equation}
E(v) = - \frac{1}{2} \eta M v^2, \;\; {\cal F}(v) = \frac{32}{5} \eta^2 v^{10} ~.
\label{EF}
\end{equation}

We use the chain rule to obtain
\begin{equation}
\frac{dE(v)}{dv} \frac{dv}{df} \frac{df}{dt} = - {\cal F}(v)
\end{equation} 
which finally results in
\begin{equation}
\frac{df}{dt} = - \frac{3 v^2}{\pi M} \frac{{\cal F}(v)}{ E'(v)}
\label{dfdt}
\end{equation}
with $E'(v) = dE/dv$.

The orbital phase is 
\begin{equation}
\frac{d\phi}{dt} = \pi f(t) ~.
\label{dphidt}
\end{equation}
The time-domain waveform is determined by integrating Eq.~\eqref{dfdt} and Eq.~\eqref{dphidt} simultaneously to determine $f(t)$ and $\phi(t)$. There are two integration constants, which are the frequency and phase at some reference time. Using the relation of the GW amplitude $h$ at a fixed distance to the GW luminosity ${\cal F} \propto (hf)^2$, the waveform can be shown to be proportional, at leading order, to
\begin{equation} \label{eq:TD_hoft}
    h(t) \propto \eta \, M^{5/3} f(t)^{2/3} \cos\left[2 \phi(t)\right] ~.
\end{equation}
As we are only interested in the time dependence, we ignore prefactors in the overall amplitude like those determined by distance and orientation, which are not relevant here.

\subsection{Frequency-domain waveform} 

Matched filtering naturally takes place in the frequency domain. We need the Fourier transform of the waveform,
\begin{equation}
\tilde{h}(f) = \int h(t) \exp(-2 \pi i f t) dt ~.
\end{equation}
The stationary phase approximation (SPA) is valid during the inspiral \cite{SPAValidity}, so we use it to compute the frequency-domain waveform. Following \cite{PoissonWill}, the SPA is
\begin{equation}
\int h(t) \, e^{i \Phi(t)} \mathrm{d}t \approx \Big( \frac{2 \pi i}{\Phi''(t_0)} \Big)^{1/2} h(t_0) \, e^{i \Phi(t_0)}
\end{equation}
with $t_0$ defined by solving for the stationary point $\Phi'(t_0) = 0$. In our case, $\Phi(t) = 2 \phi(t) - 2 \pi f\,t$ as we are computing the SPA of the Fourier transform of (Eq.~\eqref{eq:TD_hoft}). The result can be written as
\begin{equation}
\tilde{h}(f) \propto f^{-7/6} e^{i \Psi(f)} ~.
\end{equation}
As before, we ignore irrelevant prefactors. The condition on $t_0$ is $\phi'(t_0) = 2 \pi \, f$, which is the time when the gravitational-wave frequency is equal to the Fourier frequency.
The leading order amplitude of $f^{-7/6}$ is obtained because $\Phi''(t_0) = 2 \phi''(t_0) = 2 \pi (df/dt)$. The phase term is $\Psi(f) = 2 \phi(f) - 2 \pi f \, t(f)$. The gravitational-wave frequency is twice the orbital frequency, giving the factor of $2$, but in the Fourier domain the phase includes the second term which accounts for the time at which the waveform passes through that frequency. In what follows, we will calculate $\Psi(f)$ directly.

To determine the effect of the orbital resonance on the phase, we use the formulation in \cite{BIOPS}, with
\begin{equation}
\label{dtdf}
    \frac{dt}{df} =  - \frac{\pi M}{3 v^2} \frac{ E'(v)}{{\cal F}(v)} ~,
\end{equation}
which is the inverse of Eq.~\eqref{dfdt}, and
\begin{equation} 
\label{dphasedf}
    \frac{d\Psi}{df} = - 2 \pi t(f) ~.
\end{equation}

The two integration constants are now $t$ and $\Psi$ at a given frequency. Solving these two first-order differential equations gives the waveform directly in the frequency domain in the form most suitable for our goals in this paper. When a quasinormal mode is coupled to the orbit by the tidal driving, energy can be transferred between the orbit and the mode.

\section{\label{sec:resmodel}Resonance Model}

The loss of energy to the mode can be treated as an extra energy-flux $\Delta {\cal F}$ in addition to the GW energy-flux, resulting in
\begin{equation}
\label{dtdf_res}
    \frac{d(t+\Delta t)}{df} =  - \frac{\pi M}{3 v^2} \frac{ E'(v)}
{{\cal F}(v) + \Delta{\cal F}(v)} ~.
\end{equation}

In \cite{lai1994resonant}, this is instead modeled as an extra energy term, $\Delta E(D)$, where $D$ is the orbital separation. However, this energy is not directly a function of $D$, or equivalently $f$, but depends on the time history of the orbital evolution; the excitation of the mode depends on how long the tidal force dwells near the resonant frequency. So, we prefer to formulate the orbital energy loss rate to the mode as an energy-flux which is integrated in time.

\subsection{Effect of Energy Transfer}
When the energy loss to the mode excitation is small compared to the GW energy-flux, the effect on the waveform will be a small perturbation to the phase. Denoting the unmodified quantities $t_0(f)$ and $\Psi_0(f)$, and those with the resonance effect included in the dynamics as $t_{\rm res}(f)$ and $\Psi_{\rm res}(f)$, we have
\begin{eqnarray}
    \frac{d\Delta t}{df}
    &=&  \frac{d t_{\rm res}}{df} - \frac{d t_0}{df}\nonumber \\
    &=& - \frac{\pi M  E'(v)}{3 v^2} \left[ \frac{1}{{\cal F}(v) + \Delta{\cal F}(v)} - \frac{1}{{\cal F}(v)} \right] ~, \\
    \frac{d\Delta \Psi}{df}
    &=&  \frac{d \Psi_{\rm res}}{df} - \frac{d \Psi_0}{df}\nonumber \\
    &=&2\pi\Delta t~. 
\end{eqnarray}

The g-mode has long damping times, causing the resonance to be quite narrow in frequency. We can therefore approximate the process as a sharp energy transfer occurring instantaneously at a single frequency. We write the excess energy-flux as
\begin{equation}\label{eq:narrres}
    \Delta {\cal F}(f) = F_{res} ~ \delta(f - f_{\rm res})
\end{equation}
where $f_{\rm res}$ is the resonance frequency and $F_{res}$ is a proportionality factor describing the intensity of the resonance, with units of energy-flux times frequency. Energy will be transferred from the orbit to a mode when the GW frequency sweeps through the mode's resonant frequency.

\subsection{\label{sec:deltaMismatch}Phase change $\Delta\Psi(f)$ approximation for sharp resonance}

Next, we show the analytical computation of the change in the waveform when applying a sharp resonance approximation and taking the assumption $\Delta {\cal F}(v)/{\cal F}(v)\ll 1$.
In the presence of a sharp resonance, a sudden increase in GW energy-flux occurs as described in Eq.~\eqref{eq:narrres}.

Assuming that $\Delta {\cal F}(v)/{\cal F}(v)\ll 1$, we can expand Eq.~\eqref{dtdf_res} obtaining
\begin{equation}\label{approx}
\frac{d(t+\Delta t)}{df} = - \frac{M \pi E'(v)}{3 v^2 {\cal F}(v)} \left(1 - \frac{\Delta {\cal F}(v)}{{\cal F}(v)}\right)~.
\end{equation}
where the velocity can be expressed in terms of the frequency ($v=v(f)$) in the Newtonian approximation as in Eq.~\eqref{eq:v=v(f)}.

Then, we can express the time to merger difference as
\begin{equation}
\Delta t(f)= \int \frac{M \pi E'(v)}{3 v^2 {\cal F}^2(v)} \Delta {\cal F}(v) df,
\end{equation}
This can be rewritten as
\begin{equation}
\Delta t(f) = - \int \frac{dt}{df} \frac{\Delta\mathcal{F}(f)}{\mathcal{F}(v)} \mathrm{d}f
\end{equation}
where $\frac{dt}{df}$ is the time evolution without the resonance. Because $\Delta t$ is a small contribution to the evolution of $t$, we have
\begin{equation}
\int \Delta\mathcal{F} \frac{dt}{df} \mathrm{d}f =  \Delta E_{\rm res}
\end{equation}
where $\Delta E_{\rm res}$ is the total energy transferred from the orbit into the resonance when passing through $f_{\rm res}$.

The mode energy $\Delta E_{\rm res}$ accounts for the total energy transferred and can be expressed in terms of the resonant frequency and overlap integral \cite{PhysRevD.108.043003,lai1994resonant}

\begin{equation}\label{eq:dEfresQ}
    \Delta E=\Delta E(f_{\rm res},Q)~.
\end{equation}

The phase shift (cumulative change in the phase) caused by the mode excitation is a function of the mode energy, which can be written as
\begin{equation}
    \Delta \Phi=\Delta \Phi(\Delta E(f_{\rm res},Q))~.
\end{equation}

The energy transferred to the oscillatory modes is then quantified following \cite{lai1994resonant}. The lagrangian displacement of the NS fluid $\xi(t,x)$ is expanded in spherical harmonics and decomposed into a sum of oscillatory modes 
\begin{equation}
    \xi(t,x)=\sum_\alpha a_\alpha(t)\xi_\alpha(x),
\end{equation}
where $\alpha=l,m$.
The oscillatory modes fulfill the normalization  $\int dx^3\rho\xi_\alpha^*\xi_{\alpha^\prime}=\delta_{\alpha\alpha^\prime}$, where $\rho$ is the energy density. To compute the time evolution of the mode amplitude, the system is modeled as a harmonic oscillator driven by tidal force
\begin{equation}\label{eq:EOM}
    \left(\rho\frac{\partial^2}{\partial t^2}+{\cal L}\right)\xi=-\rho\nabla U~,
\end{equation}
where $U$ corresponds to the tidal field and ${\cal L}$ is an operator that accounts for the internal restoring forces of the star. This can be rewritten as 
\begin{equation}\label{amplitudes}
    \ddot{a}_\alpha+\omega_\alpha^2 a_\alpha=\frac{GM^\prime W_{lm}Q_{nl}}{D^{l+1}}e^{-im\Phi(t)},
\end{equation}
where $W_\alpha=W_\alpha(l,m)$ is a numerical coefficient depending on the mode and $Q_{nl}$ is the tidal coupling defined as
\begin{equation}
    Q_{nl}=\int d^3 x\rho\xi^*_{nlm}~·\nabla[r^l Y_{lm}(\theta,\phi)]
\end{equation}
with $r$ being the radial coordinate and $Y_{lm}(\theta,\phi)$  the standard spherical harmonic. After solving \eqref{amplitudes}, it is possible to compute the kinetic and potential energy 
\begin{eqnarray}
    E_k&=&\frac{1}{2}\sum_\alpha|\dot{a}_\alpha|^2\\
    E_p&=&\frac{1}{2}\sum_\alpha \omega^2_\alpha|a_\alpha|^2\nonumber
\end{eqnarray}

Then, the total energy transferred to the mode as a function of time can be obtained by adding up the kinetic and potential contributions
\begin{equation}\label{eq:E_tot}
    \Delta E (t)=E_k+E_p ~.
\end{equation}
This expression can be integrated to obtain  the energy shift as a function of the resonant frequency and tidal coupling from Eq.~\eqref{eq:dEfresQ}.

In the case of a narrow transferred energy-flux with the shape of a delta function as in Eq.~\eqref{eq:narrres} the integral collapses to:
\begin{equation}\label{eq:dt-sharp}
\Delta t(f) = \Delta t_{\rm res}~\Theta(f-f_{\rm res}),
\end{equation}
with
\begin{equation}
\Delta t_{\rm res} = - \frac{\Delta E_{\rm res}}{\mathcal{F}(v_{\rm res})}
\end{equation}
and where the Heaviside function is $\Theta(f-f_{\rm res}) = 1$ for $f \geq f_{\rm res}$ and $\Theta(f-f_{\rm res}) = 0$ for $f<f_{\rm res}$.
 
The change in phase can similarly be computed integrating Eq.~\eqref{dphasedf}
\begin{equation}\label{eq:dpsi-sharp}
\Delta\Psi(f) = 2 \pi (f-f_{\rm res}) \Delta t_{\rm res}~\Theta(f-f_{\rm res}) ~.
\end{equation}

\begin{figure}[H]
\includegraphics[width=\linewidth]{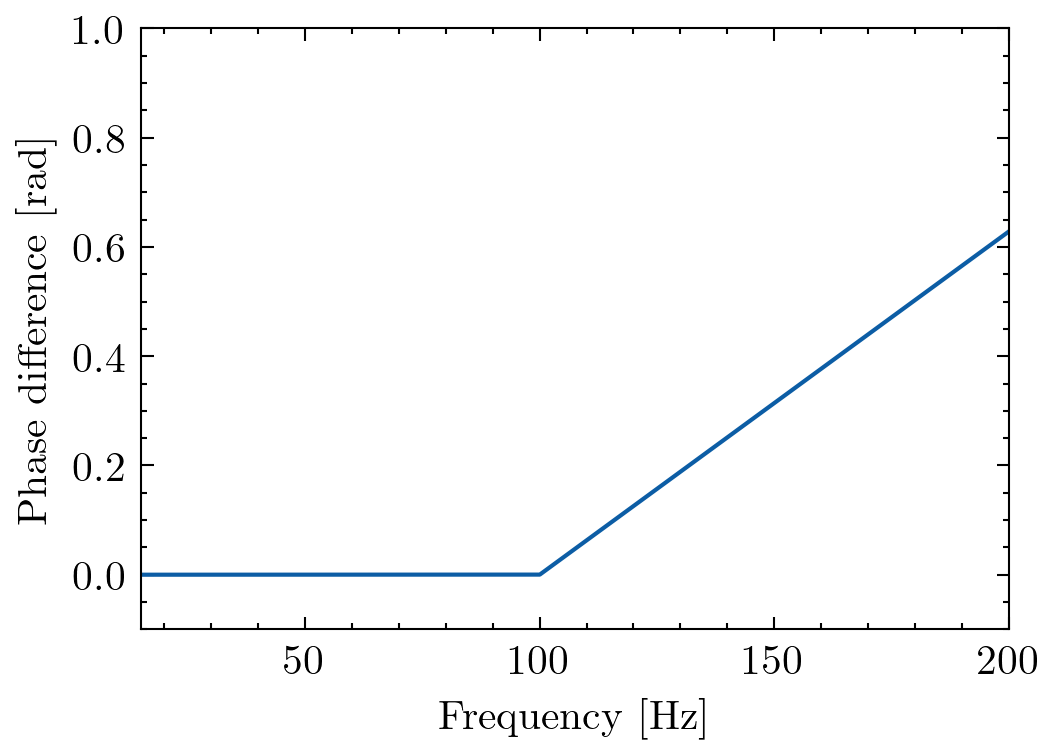}
\caption{\label{fig:0_basis/dpsi} Phase difference computed using the sharp resonance approximation from Eq.~\eqref{eq:dpsi-sharp}.}
\end{figure}

Fig.~\ref{fig:0_basis/dpsi} shows the phase difference as a function of frequency for a time shift of $1~ms$ at a resonant frequency of $100~Hz$.

\subsection{\label{sec:ede}Phase jumps in the context of NS resonance}

For the excitation of resonant modes in the neutron star to occur, a portion of the orbital energy of the binary system must be transferred to the stellar modes. Assuming that no energy is transferred back from the modes to the orbit, we obtain a preferred direction condition for the energy transfer:
\begin{equation}\label{eq:dir}
\Delta{\cal F}(f) \ge 0~.
\end{equation}

In a more realistic scenario, small oscillations in the transferred energy flux are expected after the resonance. However, this does not occur in the idealized sharp resonance case considered here. We will adopt this condition to analyze the consistency of a phase jump in the waveform caused by resonance effects, as modeled in \cite{Edelman_2021}.

Using Eq.~\eqref{dphasedf}, the second derivative of the frequency-domain phase difference with respect to frequency can be written as
\begin{eqnarray}
\frac{d^2\Delta\Psi}{df^2} &=&2\pi\frac{d\Delta t}{df} 
\\
&=& -\frac{2\pi^2 M E^\prime(v)}{3 v^2 {\cal F}(v)} \frac{\Delta {\cal F}(v)/{\cal F}(v)}{1 + \Delta{\cal F}(v)/{\cal F}(v)} ~,\nonumber
\end{eqnarray}

where $E^\prime(v)<0$. The term $\frac{\Delta {\cal F}(v)/{\cal F}(v)}{1 + \Delta{\cal F}(v)/{\cal F}(v)}$ is  positive when Eq.~\eqref{eq:dir} is satisfied. Thus, if $\Delta{\cal F}(v)\ge 0$, the curvature of the phase difference remains positive, implying that
\begin{equation}
    \frac{d^2\Delta\Psi}{df^2}\ge 0~.
\end{equation} 

This condition is not fulfilled in the case studied in \cite{Edelman_2021}, where the phase difference is modeled using a step function and a hyperbolic tangent. The curvature of these functions is not positive through all its domain, which violates the assumption $\Delta{\cal F} \ge 0$. In fact, to obtain a step function in $\Psi(f)$ would require a sudden large positive value of $\Delta\mathcal{F}$ closely followed by an equally large negative value; this is likely unphysical.

In case of relaxation of the sharp resonance approach to recover a more realistic scenario, like the one proposed in \cite{lai1994resonant}, small oscillations of the energy flux around resonance are admitted. Then, the statements above could be rewritten as $\Delta{\cal F} \gtrsim 0$.

\section{Methods for Estimating the Detectability of Resonance Effects}\label{sec:dettech}

While the methodology developed in Sec. \ref{QA}  is valid for the detection of any astrophysical signal where the true time and phase are unknown, here it is applied to detecting the subtle time advance resulting from BNS resonance. The challenge is to resolve the effect, which can be done through (1) increasing resolution and actively maximizing the overlap function to determine the match between the perturbed and unperturbed waveforms or (2) expanding the match function up to quadratic order following the metric approximation of Owen \& Sathyaprakash 1998 \cite{owen1999matched}. The methods agree for small phase changes/time advances with the metric approximation being faster. In addition to the more rapid computation, the expansion in moment integrals has the advantage of providing a method that agrees with the FFT method but is a much simpler computation.

For definiteness, we consider a binary inspiral formed from two neutron stars with masses $m_1=m_2=1.4 \ M_\odot$. We assume that neither neutron star is spinning. The waveform is modeled via the IMRPhenomD approximant with its lower frequency bound fixed at $15~ {\rm Hz}$. Unless stated otherwise, its high frequency bound at $1024~ {\rm Hz}$, the sampling rate is fixed at $4096~ {\rm Hz}$ and the duration of the matched filter time series at $512~ {\rm s}$. The noise curve is an analytical model similar to the LIGO O4 noise curve \cite{PyCBC_aLIGO175MpcT1800545}.


\subsection{Comparison of Match Computation Methods}

The match (Eq.~\eqref{eq:match_definition}) is closely related to the SNR timeseries (Eq.~\eqref{eq:SNR_ts}) when as the data we take a normalized waveform, i.e. $d(f) = \frac{b(f)}{\sqrt{\langle b, b \rangle}}$ and as the template $h(f) = a(f)$. The match is the maximum over phase and time of the resulting timeseries. Clearly the maximum possible value is one. This can be calculated efficiently using the Fast Fourier transform.
We use the PyCBC software package \cite{pyCBC,Allen:2005fk,Nitz:2017svb,DalCanton:2014hxh}
to perform the match computation.
In the standard method the time series is only calculated on a grid of times spaced at the sample rate of the FFT. This method can miss the true peak by a time shift of up to half the distance between time samples. When trying to calculate a small mismatch, this inaccuracy may be significant.
There is a subsample interpolation method implemented in PyCBC which fits a quadratic function to three SNR values around the maximum to improve the resolution of the location and value of the maximum. However, the accuracy of this estimate is not guaranteed.
Another method is to calculate the SNR at only one time offset, and then use numerical optimization to iteratively find the peak. This is available in PyCBC as the \texttt{pycbc.filter.optimized\_match} function.

A final method to calculate the match is the quadratic approximation (Eq.~\eqref{eq:match_quad}). This only requires calculating the several moment integrals which compose (Eq.~\eqref{eq:quadratic_coefficient}) to obtain a single coefficient. Since it is derived from a Taylor series, the approximation will fail past some value of $\epsilon$, but in all practical uses we are only concerned with matches close to one where the approximation is very accurate. This method can also be easily extended to account for other nuisance variables like the chirp mass.

\begin{figure}[h]
\includegraphics[width=\linewidth]{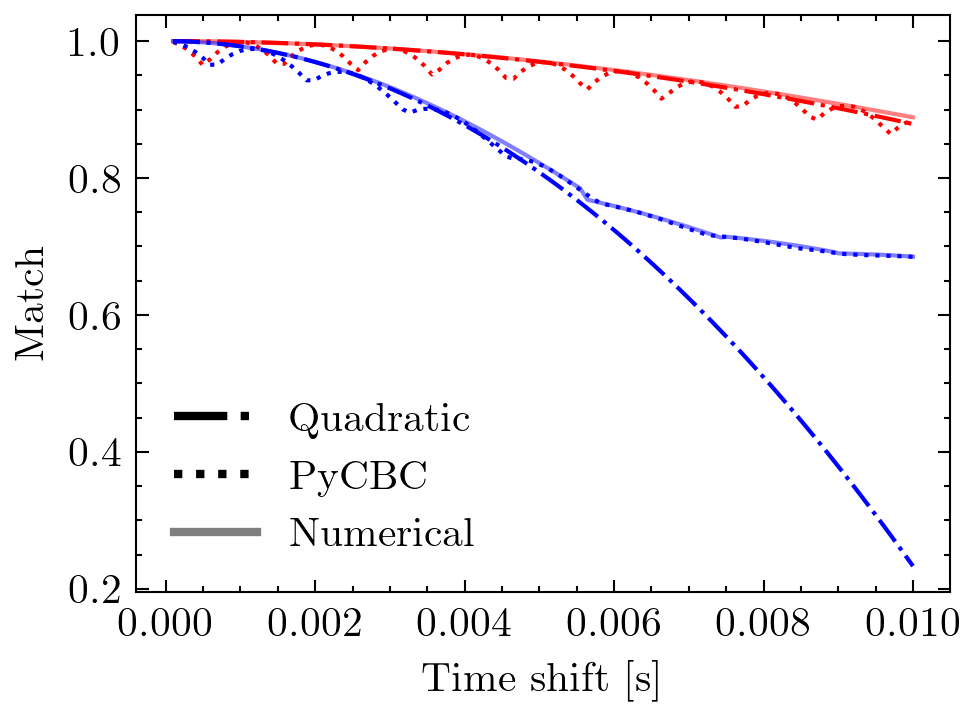}
\includegraphics[width=\linewidth]{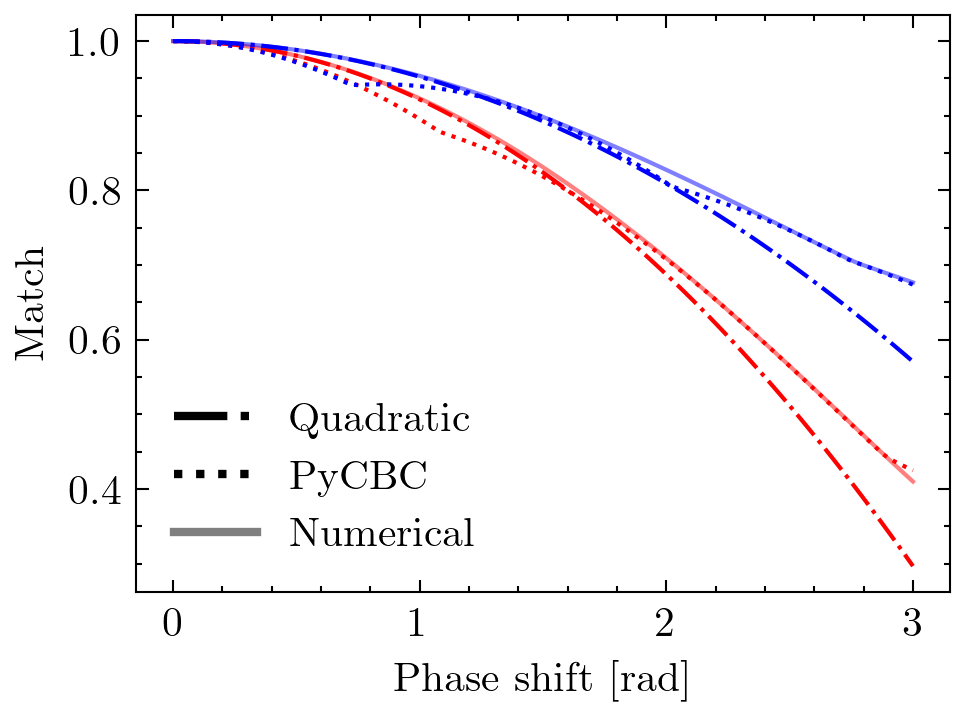}
\caption{\label{fig:I} Match computation applying the quadratic approximation from Eq. \eqref{eq:match_quad} (dot-dashed curves), using PyCBC match function \texttt{pycbc.filter.match} (dotted curves) and the optimized numerical computation (fading continuous curves). In the plot above, we express the match in terms of the time shift and in the one below in terms of the phase shift at different resonant frequencies ($50~ {\rm Hz}$ in red and $100~ {\rm Hz}$ in blue).}
\end{figure}

We compare these three methods for match computation in Fig.~\ref{fig:I}. For these figures, we used a sampling rate of $1024~ {\rm Hz}$ and a higher frequency bound of $512~ {\rm Hz}$. The time shift from Eq.~\eqref{eq:dt-sharp} and the phase shift from Eq.~\eqref{eq:dpsi-sharp} are parameterized by $\Delta t$ and $\Delta\psi$, respectively, quantifying the intensity of the perturbation. These variables play the role of $\epsilon$ in their respective quadratic approximations. Note that these are the time / phase shift of the inspiral after the resonance frequency relative to before the resonance. They are not the same as the overall $\delta t$ and $\delta\phi$ which affect the entire waveform and which are maximized over in the match computation.

In both Fig.~\ref{fig:I} (a) and (b) the FFT-based match calculation exhibits oscillations due to the finite time sampling.  The accuracy can be improved by increasing the sampling rate. On the other hand, the results from the quadratic approximation are smooth and the code is much faster. The quadratic approximation agrees with the numerically optimized value for small values of the mismatch, but for larger values the quadratic approximation starts to overestimate the mismatch. 

\subsection{Intuition for SNR}

To understand the behavior of the match for large $\epsilon$, we can split the SNR computation into two parts: the part arising before the resonance and the part after the resonance. The template is the BNS waveform with no resonance, $h_0(f)$. Then the SNR timeseries for any signal waveform $h_s(f)$ is
\begin{equation}\label{eq:SNR_t_phi}
    \rho(t, \phi; h_s)
    = \frac{4}{|h_0|} \mathrm{Re} \left[ \int_{f_{\rm min}}^ {f_{\rm max}} \mathrm{d}f \frac{h_0^*(f) h_s(f)}{S_n(f)} e^{-i (2 \pi f \, t + \phi)} \right]~ ,
\end{equation}
with $|h_0| \equiv \sqrt{\langle h_0, h_0 \rangle}$.
We split the resonant waveform into a sum $h_\epsilon(f) = h_a(f) + h_b(f)$ of the part of the frequency-domain waveform before ($h_a$) and after ($h_b$) the resonance in frequency:
\begin{eqnarray}\label{eq:snr_split}
    h_a(f) &=& h_\epsilon(f) \, (1 - \Theta(f - f_{\rm res})) \\
    h_b(f) &=& h_\epsilon(f) \, \Theta(f - f_{\rm res}) \nonumber ~.
\end{eqnarray}
By linearity, the full SNR timeseries is the sum of the SNRs from these two pieces, which we will call $\rho_a(t)$ and $\rho_b(t)$. We maximize over the phase (replacing $\mathrm{Re}$ with $\mathrm{Abs}$ in (Eq.~\eqref{eq:SNR_t_phi})) in order to keep the plots legible.

It is also possible to apply the quadratic approximation from Sec.~\ref{QA} to the SNR time series from Eq.~\eqref{eq:SNR_ts}. Maximizing over the phase but leaving the time variable intact gives
\begin{equation}\label{eq:quadsnr}
    \rho_{q}(t)=1 - \frac{1}{2} \left( \mathcal{J}[(\Delta\Psi+2\pi f t)^2] - \mathcal{J}[\Delta\Psi+2\pi f t]^2 \right)~.
\end{equation}
This can be compared with the absolute value of $\rho(t)$.

\begin{figure}[h]
\includegraphics[width=\linewidth]{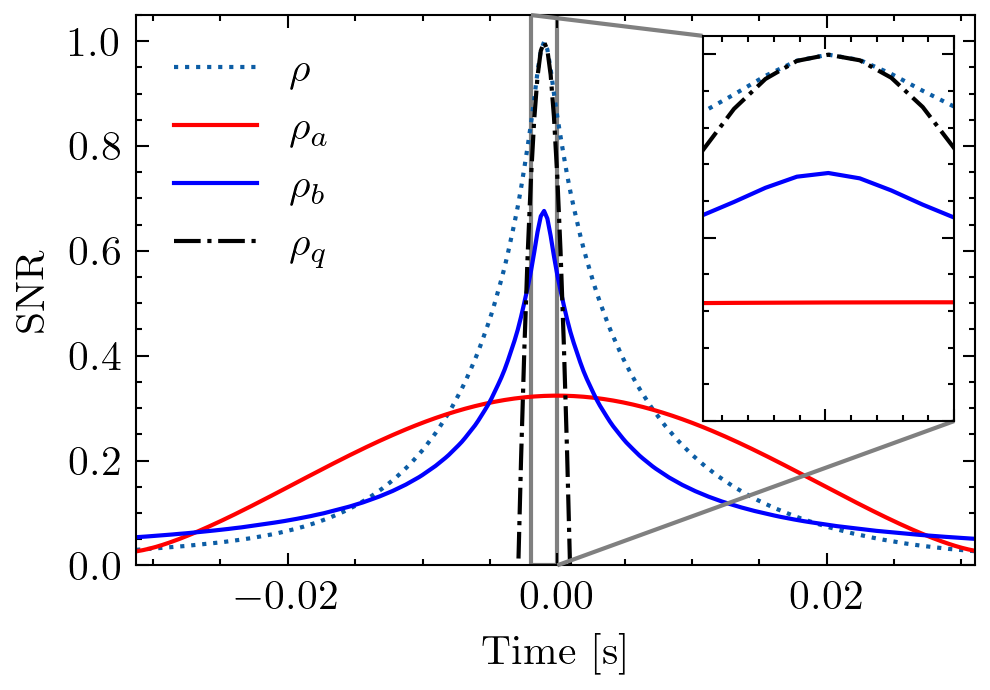}
\includegraphics[width=\linewidth]{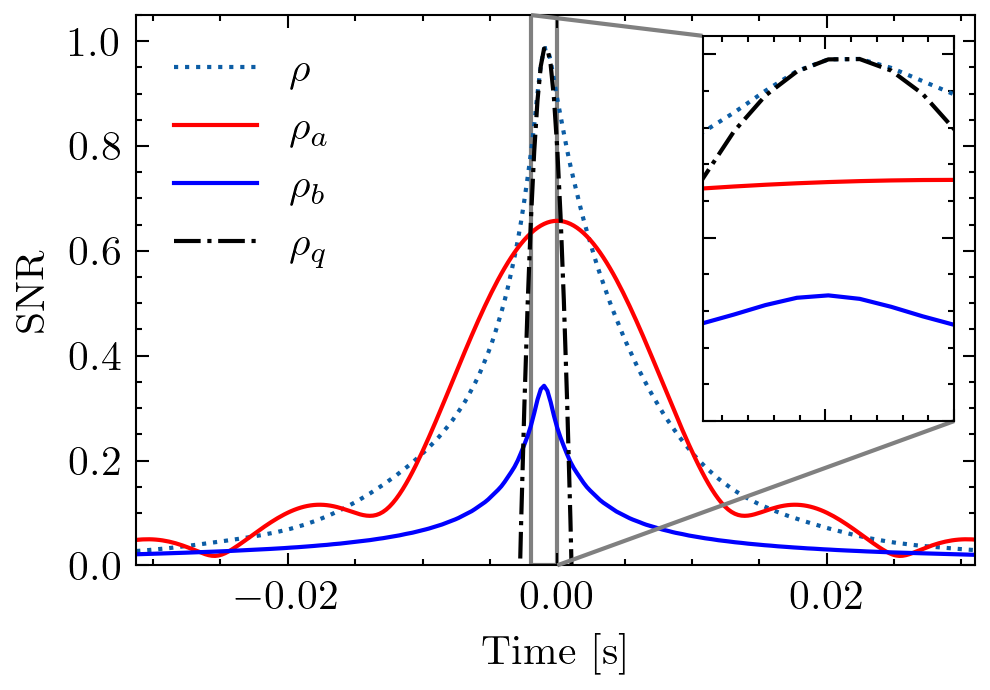}
\caption{\label{fig:rho_abs_roll_split_1ms} Dotted blue curve: Absolute value of the normalized SNR amplitude (Eq.~\eqref{eq:SNR_t_phi} maximized over the phase $\phi$). Red and blue solid curves: contributions to the overlap from the waveform segments before and after the resonance, respectively, from \textbf{Eq.~\eqref{eq:snr_split}}. The dot-dashed curve corresponds to the quadratic approximation of the SNR in Eq.~\eqref{eq:quadsnr}. The time shift is fixed to $1~{\rm ms}$ and the resonant frequencies from each plot are $50~{\rm Hz}$ (above) and $100~{\rm Hz}$ (below).}
\end{figure}

\begin{figure}[h]
\includegraphics[width=\linewidth]{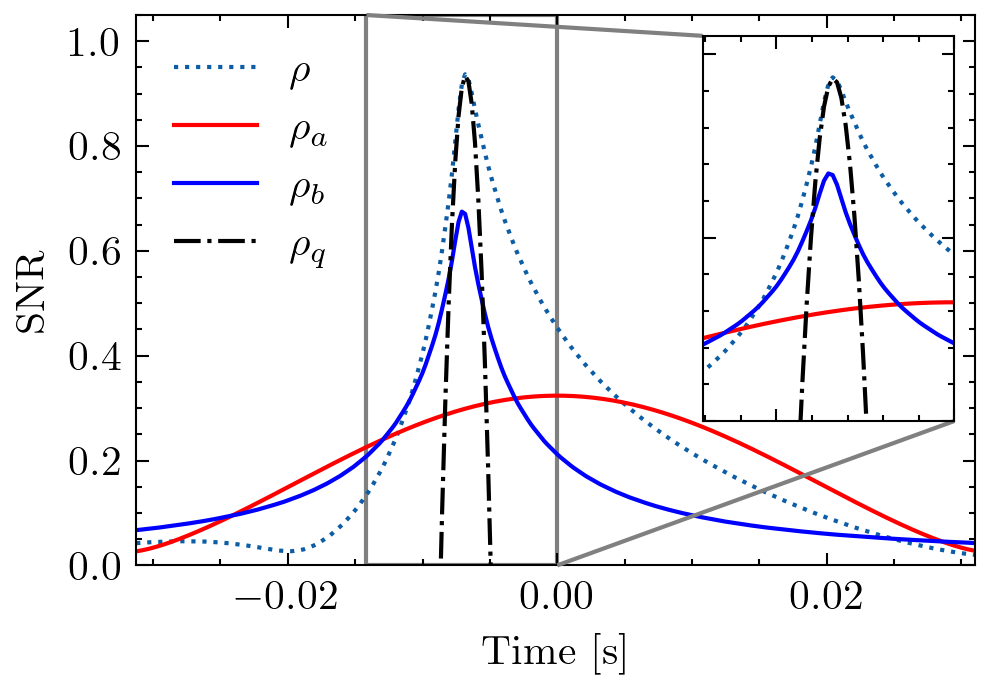}
\includegraphics[width=\linewidth]{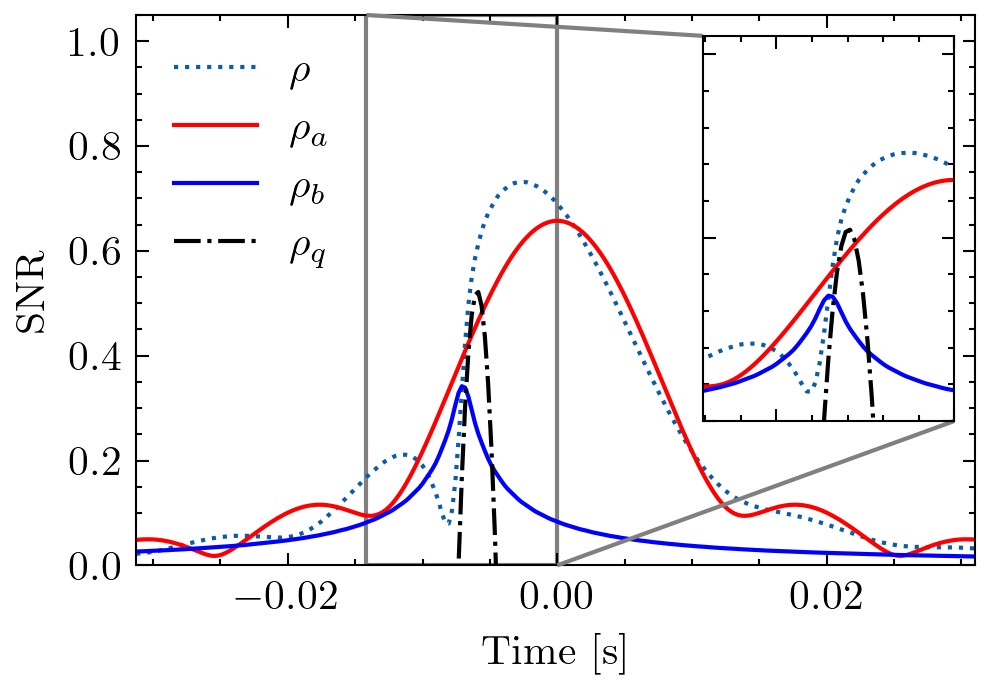}
\caption{\label{fig:rho_abs_roll_split_7ms} Same as Fig.~\ref{fig:rho_abs_roll_split_1ms} but changing the time shift to $7~ {\rm ms}$.}
\end{figure}

Fig.~\ref{fig:rho_abs_roll_split_1ms} and Fig.~\ref{fig:rho_abs_roll_split_7ms} show the SNR time series for a time shift of $\Delta t = 1~{\rm ms}$ and $\Delta t =7~{\rm ms}$ at two resonant frequencies $f_{\rm res} = 50$ Hz and $f_{\rm res} = 100$ Hz, respectively.
The SNR curves are normalized by ensuring that $\langle h_\epsilon, h_\epsilon \rangle = 1$.
The time shift due to the resonance shifts the peak of $\rho_b$ relative to $\rho_a$. For small time shifts, the two peaks nearly overlap and so the quadratic approximation works well; it even accounts for the phase, which our plots do not capture. However, for larger time shifts (due to a stronger resonance effect), the two peaks separate. The template $h_0$ can be aligned in time with either $h_a$ or $h_b$, but not both, and when it is aligned with one the SNR timeseries of the other is small. Hence as the time shift increases, the match will asymptote to the larger of the peak values of $\rho_a$ or $\rho_b$.

\subsection{Connection of "indistinguishability" criterion and match}

An effect is detectable when it induces a change in the waveform which can be distinguished from a noise fluctuation. A commonly used criterion for indistinguishability of two waveforms \cite{Lindblom_2008} is the inequality
\begin{equation}\label{mmlindblomres}
    \langle\delta h,\delta h\rangle < 1
\end{equation}
where $\delta h = h_1-h_0$. Here $h_1$ and $h_0$ are the waveform with and without the effect in question present. This criterion requires that the difference in waveforms be smaller than even a very small noise fluctuation of one standard deviation. 

Turning this around, the criterion for an effect to be \emph{detectable} is $\langle\delta h,\delta h\rangle > 1$. For a more stringent criterion of $n_\sigma$ standard deviations, we require
\begin{equation}\label{eq:lindblom_sigma}
    \langle\delta h,\delta h\rangle > n_\sigma^2 ~.
\end{equation}
This means that the estimate of the size of the effect would be $n_\sigma$ standard deviations away from zero.

To relate the match to this criterion, we expand
\begin{eqnarray}
\langle \delta h, \delta h\rangle &=& \langle h_1 - h_0, h_1 - h_0 \rangle \nonumber\\
&=& \langle h_1, h_1 \rangle + \langle h_0, h_0 \rangle - 2 ~\langle h_1, h_0 \rangle ~.
\end{eqnarray}
Writing the SNR of the signal $|h_1| \equiv \rho$, and noting that $|h_0| = |h_1|$ because they differ only by a phase term, we have
\begin{equation}
\langle \delta h, \delta h\rangle = 2 \rho^2 (1 - {\cal M}(h_1, h_0) ~) ~.
\end{equation}

The detectability threshold is given by
\begin{equation}
\label{eq:mismatch_criterion}
1-{\cal M}(\Delta\Psi ) > \frac{n_\sigma^2}{2\rho^2} ~.
\end{equation}


\section{Detectability of Resonance Effects}\label{sec:detresults}

As in the previous section, we consider a standard NS-NS binary inspiral formed from two neutron stars with masses $m_1=m_2=1.4 \ M_\odot$ and without spins. When comparing different detectors, we place the binary at a luminosity distance of $d_L=100~ {\rm Mpc}$ and with optimal orientation. When comparing results at different SNRs, the distance is varied to obtain the desired SNR.

\subsection{O5 Detectability Prospects}

\begin{figure}[h]
\includegraphics[width=\linewidth]{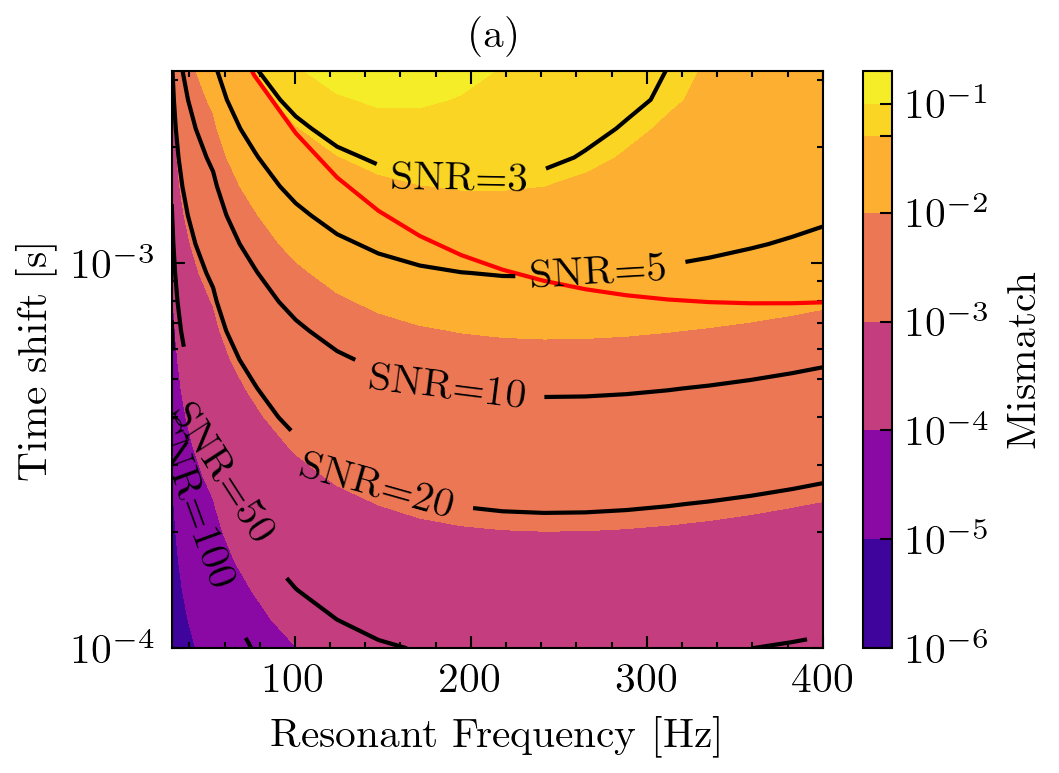}
\includegraphics[width=\linewidth]{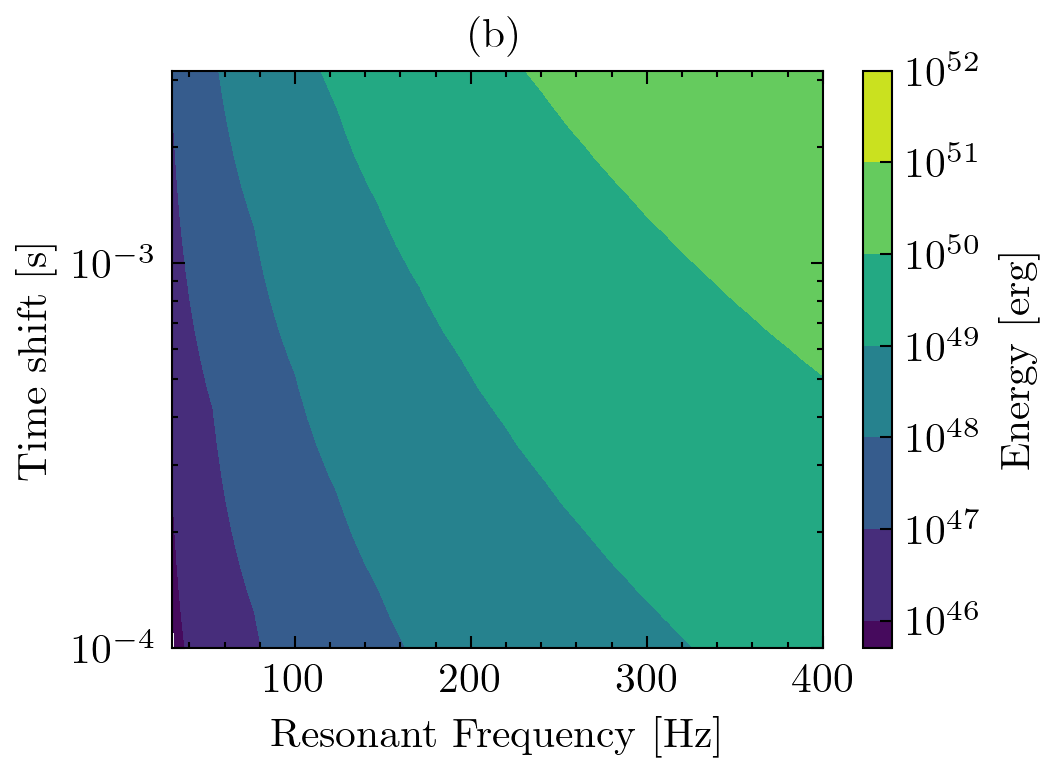}
\caption{\label{fig:delta_mismatch} (a) Mismatch as a function of the time shift $\Delta t$ and the resonance frequency $f_{\rm res}$. Black curves indicate the signal-to-noise ratio (SNR) required for detecting a time shift of a given value, satisfying the Lindblom criterion in Eq.~\eqref{mmlindblomres}. The red curve marks the transition between using the refined numerical match function and the quadratic approximation 
for a relative error among the match functions of $0.1 \%$.
(b) Total energy transferred to the neutron star modes, obtained integrating Eq.~\eqref{dtdf_res}.}
\end{figure}

We now investigate the detectability of a resonance with the currently anticipated detector noise curve for O5 (the A+ design PSD) \cite{Barsotti2018Aplus}. The mismatch is a function of both the frequency of the resonance and the time shift $\Delta t$ of the merger caused by passing through the resonance. Fig.~\ref{fig:delta_mismatch} shows both the mismatch and the SNR required for detection at $1~\sigma$ according to the criterion in Eq.~\eqref{eq:mismatch_criterion}. The red line marks the transition from the region where the quadratic approximation is valid (below the curve) to where the match must be computed numerically, with validity defined by a relative error of $0.1 \%$.

The parameters of the resonance can also be mapped to the energy or energy-flux.
A mode excitation that leads to a $\Delta t = 1 ~ \text{ms}$ at $f_{\rm res} = 100~ {\rm Hz}$ corresponds to a mismatch of about 1\%, which is within the detectability threshold for O5 at an SNR of $8$, where the energy transferred to the mode is $\sim 10^{48}~ {\rm erg}$ and the flux fraction is $\sim 1\%$. We note that internal neutron star physics can significantly affect the resonance frequency. In particular, the presence of leptonic buoyancy may raise the frequency of a $\ell=m=2$ g-mode to values of the order of $f_{\rm res} \sim 400~ {\rm Hz}$ \cite{Rau_2018}. Estimating these effects is beyond the scope of this first paper.

\subsection{Detector comparison}

\begin{figure}[H]
\includegraphics[width=\linewidth]{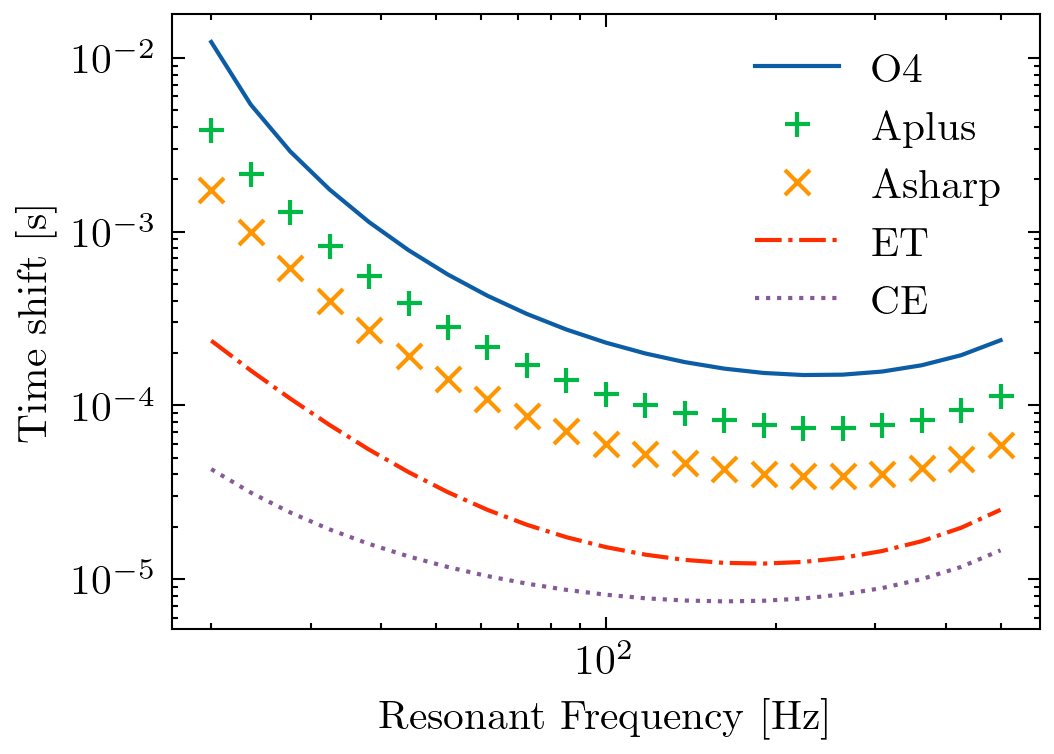}
\caption[Detectability thresholds]{
Merger time advance detectability threshold for different detectors as computed using the numerical match calculation.
\label{fig:MDet}}
\end{figure}

Finally, in Fig.~\ref{fig:MDet} we compare the detectable time advance shown as a function of $f_{\rm res}$ for different detectors. The noise curves used include O4 \cite{PyCBC_aLIGO175MpcT1800545} followed by A+ (O5) \cite{LIGO_AplusDesign_T2000012} and A\# (post O5) \cite{LIGO_AsharpStrain_T2300041}. The proposed ET \cite{ET} and CE \cite{CE} noise curves come with a significant sensitivity improvement of almost two orders of magnitude beyond current detectors. The low frequency threshold fixed for each detector is: $15~{\rm Hz}$ for O4, $10~{\rm Hz}$ for A+ and A\# and $6~{\rm Hz}$ for ET and CE.

\subsection{Comparison with previous approximations}

A commonly used approximation for detectability is given in \cite{read2023waveform}. In our notation, it reads
\begin{eqnarray} \label{eq:read_integral}
\langle \delta h, \delta h \rangle
&=& 4 \int \frac{|h_0(f)|^2 | 1 - e^{i \, \delta\phi}|^2}{S_n(f)} df
\nonumber \\
&=& 8 \int \frac{|h_0(f)|^2 ( 1 - \cos \delta\phi)}{S_n(f)} df~.
\end{eqnarray}
 We correct a missing factor of $4$ in the second equality in \cite{read2023waveform}, which however does not effect the final result.
Here, $\delta\phi$ is equivalent to our $\Delta\Psi(f)$ except with a weighted least-squares fit of the form $\phi_0 + 2\pi t_0 f$ removed, with weights $|h_0(f)|^2/S_n(f)$. This scheme is equivalent to the maximization of the match over time and phase.

For small $\delta\phi$, $2 ( 1 - \cos \delta\phi) \approx \delta\phi^2$.
We can recast  (Eq.~\eqref{eq:read_integral}) as 
\begin{equation} \label{eq:read_approximation}
\langle \delta h, \delta h \rangle
\approx \int \frac{ 4 f ~ |h_0(f)|^2 (\delta\phi)^2}{S_n(f)} d (\log f) ~.
\end{equation}
If we then assume that $|\delta \phi (f)| < \frac{\sqrt{S_n(f)}}{2|h_0(f)|\sqrt{f}}$ at all frequencies, then we would have $\langle \delta h, \delta h \rangle < \int d (\log f)$. This integral is of order one when taken over the sensitive frequency range of the detector, and so this inequality is a rough approximation for indistinguishability of waveforms.

In \cite{Counsell:2024pua}, this is turned around into a result for distinguishability which we will call the "single-frequency approximation":
\begin{equation}\label{read_phi}
    |\delta \phi (f)| > \frac{\sqrt{S_n(f)}}{2|h_0(f)|\sqrt{f}}~.
\end{equation}
Here $f$ is taken as the frequency at which the resonance occurs. This expression is quite inaccurate when used to determine detectability because the match integral is not well approximated by the value at any one frequency. The mismatch really depends on the integrated dephasing across the frequency range, weighted by sensitivity. The moment integrals used in our quadratic approximation capture the needed information much more effectively and are not very difficult to compute.

In \cite{Counsell:2024pua}, the single-frequency approximation is used to predict detectability of a BNS resonance. In that paper and others \cite{ho2023new, lai1994resonant}, the effect is treated as if it were a phase shift at the resonant frequency rather than a time shift, i.e. $\Delta\Psi(f) = \delta\phi \, \Theta(f-f_{\rm res})$ instead of our (Eq.~\eqref{eq:dpsi-sharp}), with $\delta\phi \equiv 2 \pi f_{\rm res} \Delta t_{\rm res}$. There is also no least-squares fit performed to maximize over time and phase.

\begin{figure}[h]
\includegraphics[width=\linewidth]{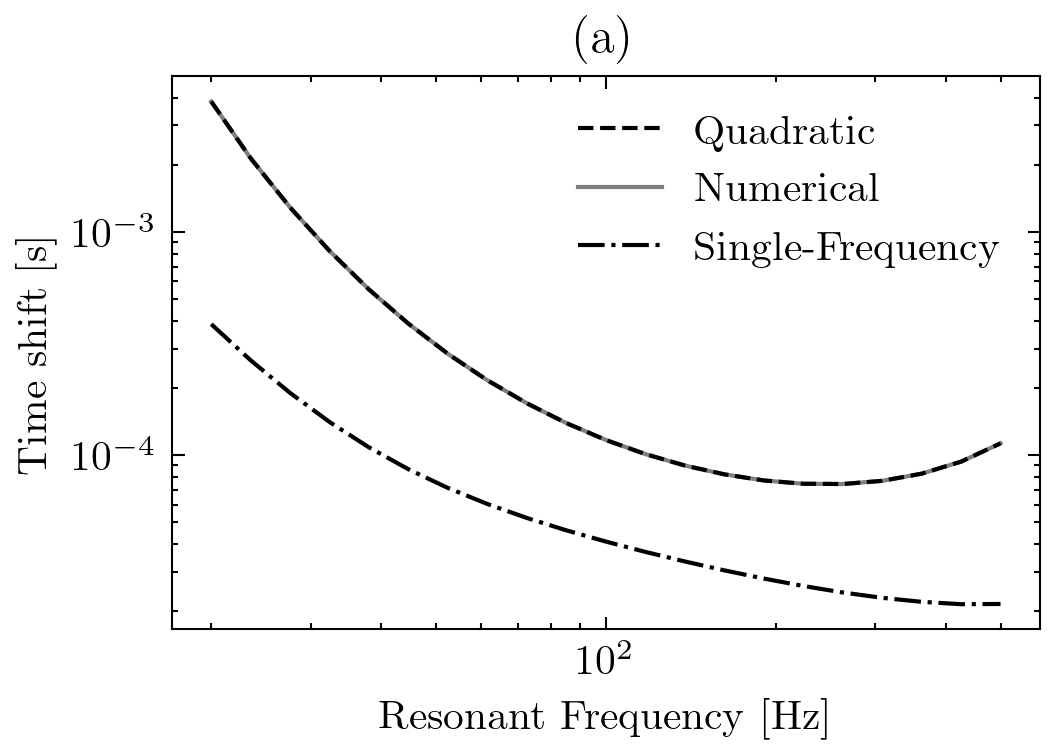}
\includegraphics[width=\linewidth]{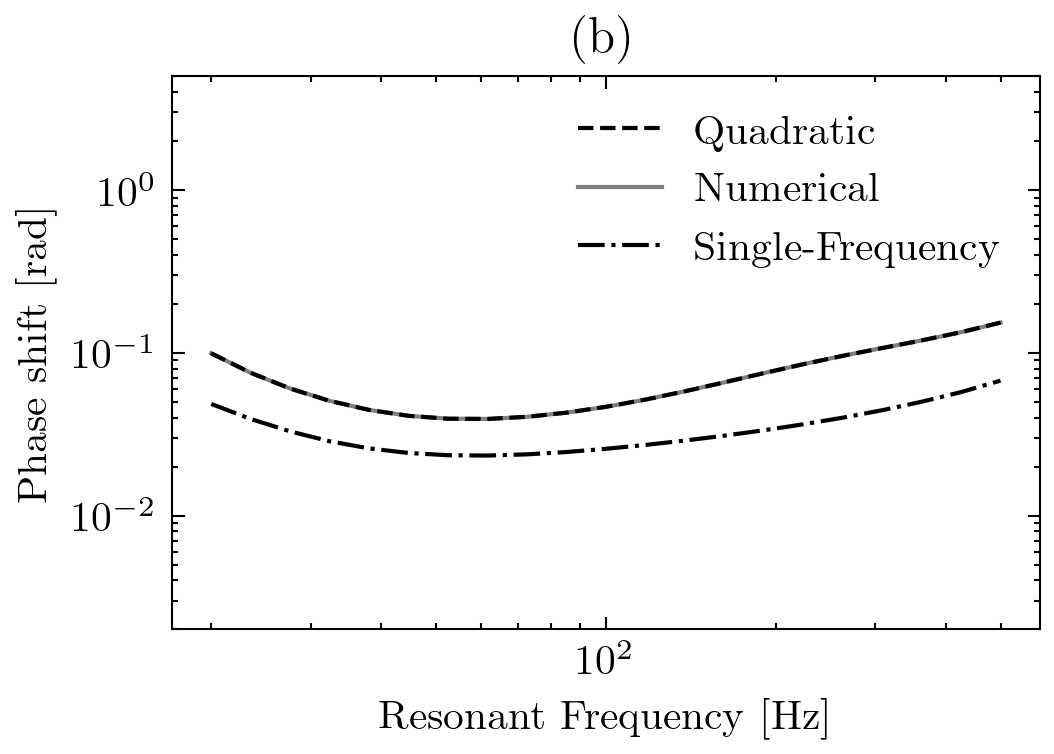}
\caption[Detectability thresholds]{(a) Time and (b) Phase shift detectability thresholds using A+ noise curves. The dashed lines correspond to the quadratic approximation. The fading continuous lines represent the numerical match calculation, and dot-dashed lines are generated by adapting Eq.~\eqref{read_phi} from \cite{read2023waveform}. The single-frequency approximation from \cite{read2023waveform}  significantly underestimates the minimum detectable values}. 
\label{fig:III}
\end{figure}

Fig.~\ref{fig:III} shows the threshold values for detectable time and phase shifts, respectively, for different detector sensitivities. To obtain this figure, we increase the sample rate to $16384 ~{\rm Hz}$ and fix the low frequency threshold to $10~{\rm Hz}$. The numerical values used satisfy the Lindblom detectability criterion from Eq.~\eqref{mmlindblomres}. As expected, since the detectable shifts are small, the quadratic approximation and the refined numerical match functions coincide. 
A key result is that the single-frequency approximation as applied by \cite{Counsell:2024pua} substantially overestimates the detectability of the resonance effect. This disagreement is not unexpected since \cite{read2023waveform} approximates an integral using the integrand value at one fixed value of $f$, while we either compute the integral using the pyCBC match function or approximate it accurately using moment integrals in the already tested quadratic approximation, both yielding overlapping results.

\section{\label{sec:concl} Conclusions}
In this paper, we consider perturbations that affect the waveform only through a frequency-dependent phase factor, without affecting the amplitude. Phase changes in the gravitational wave are the next frontier in gravitational wave astrophysics, which is precision science that constrains known physics beyond the laboratory from neutron star resonance effects to alternative theories of gravity. Our paper revises the formalism required for the detection of changes in the phase of the waveform with respect to some waveform template. We apply this formalism to determine the detectability of a resonant mode which is excited during the inspiral of a binary neutron star system. The transfer of energy from the orbit into the vibrational mode causes the merger time to be shifted earlier, but only in the part of the waveform at frequencies above the resonant frequency.

Our results agree with the numerically-computed match for both phase shifts of the gravitational wave signal in the frequency-domain and shifts in the time to merger. By comparison, the calculation proposed in \cite{read2023waveform} provides an estimate of the detectability which is a factor of several too optimistic.

This approximation is used to estimate the amount of time advance that must be produced by a BNS resonance in order to be detectable.  The energy transfer occurs when the gravitational wave frequency sweeps through the resonant frequency of a mode as a result of the coupling among the NS eigenmode and the tidal field of the companion. Correct identification of resonant modes, and the shape and duration of the resonance can constrain interior neutron star physics. Moreover, when the energy transferred is sufficient to shatter the neutron star crust it can result in coincident gravitational wave and electromagnetic emission.

In this paper the resonance is assumed to be sudden inducing a sharp change in the gravitational wave energy-flux, and a wide range of possible resonant frequencies and amplitude of resonance are explored. The next step is to consider a detailed model of the vibrational modes and the impact of the equation of state on the shape and duration of the resonance.
\section{Data availability}
The code used to generate the figures in this manuscript is publicly available at {\tt https://github.com/albertorevillap/gwsense/}.

\begin{acknowledgments}
We acknowledge the ICCUB Virgo group and the LIGO–Virgo Collaboration’s Extreme Matter group for their support and scientific exchange. In particular, we thank Arnau Rios, Roque Marquez Rodriquez, Jose Antonio Font, Pablo Cerd\'a Dur\'an, Vanessa Graber, Nils Andersson, Ira Wasserman, David Tsang, Duncan Neill, Tanja Hinderer, and François Foucart for many insightful discussions, and Marta Colleoni for helpful comments on the manuscript. We are also grateful to the ICCUB administrative team including Ana Argudo and Esther Pallarés, for their invaluable assistance.

This material is based upon work supported by NSF's LIGO Laboratory which is a major facility fully funded by the National Science Foundation. We are grateful to Marta Colleoni and Pablo Cerd\'a Dur\'an  for their internal LIGO and Virgo pre-publication reviews.

The authors acknowledge financial support from the grants RYC2022-035983-I, PID2021-125485NB-C22, PID2022-137268NB-C52
, PID2024-159689NB-C22
, CEX2019-000918- M and CEX2024-001451-M funded by MCIN/AEI/10.13039/501100011033 (State Agency for research of the Spanish Ministry of Science and Innovation) and SGR-2021-01069 (AGAUR).

\end{acknowledgments}

\bibliographystyle{apsrev4-1}
\bibliography{References.bib}

\end{document}